# The extremely large magnetoresistance in the Candidate Type-II Weyl semimetal MoTe$_2$

F. C. Chen[1,2†], H. Y. Lv[1†], X. Luo[1*], W. J. Lu[1*], Q. L. Pei[1], G. T. Lin[1,2], Y. Y. Han[3], X. B. Zhu[1], W. H. Song[1], and Y. P. Sun[3,1,4*]

[1] Key Laboratory of Materials Physics, Institute of Solid State Physics, Chinese Academy of Sciences, Hefei, 230031, China

[2] University of Science and Technology of China, Hefei, 230026, China

[3] High Magnetic Field Laboratory, Chinese Academy of Sciences, Hefei, 230031, China

[4] Collaborative Innovation Center of Advanced Microstructures, Nanjing University, Nanjing, 210093, China

## Abstract

We performed the angle dependent magnetoresistance (MR), Hall effect measurements, the temperature dependent magneto-thermoelectric power (TEP) $S(T)$ measurements, and the first-principles calculations to study the electronic properties of orthorhombic phase MoTe$_2$ ($T_d$-MoTe$_2$), which was proposed to be electronically two-dimensional (2D). There are some interesting findings about $T_d$-MoTe$_2$: (1) A scaling approach $\varepsilon_\theta=(sin^2\theta+\gamma^{-2}cos^2\theta)^{1/2}$ is applied, where $\theta$ is the magnetic field angle with respect to the $c$ axis of the crystal and $\gamma$ is the mass anisotropy. Unexpectedly, the electronically 3D character with $\gamma$ as low as 1.9 is observed in $T_d$-MoTe$_2$; (2) The possible Lifshitz transition and the following electronic structure change can be verified around $T$~150 K and $T$~60 K, which is supported by the evidence of the slop changing of the temperature dependence of TEP, the carrier density extracted from Hall resistivity and the onset temperature of $\gamma$ obtained from the MR measurements. The extremely large MR effect in $T_d$-MoTe$_2$ could originate from the combination of the electron-hole compensation and a particular orbital texture on the electron pocket, which is supported by the calculations of electronic structure. Our results may provide a general scaling relation for the anisotropic MR and help to recognize the origins of the MR effect in other systems, such as the Weyl semimetals and the Dirac ones.

†These authors contributed equally to this work.

Corresponding author: xluo@issp.ac.cn, wjlu@issp.ac.cn and ypsun@issp.ac.cn.

Materials with large magnetoresistance (MR), where their resistance change significantly with an applied magnetic field, are a driving ingredient in modern electronic devices such as hard drives in computers. In general, the MR in the simple metal is very small and saturated under high magnetic field. The discoveries of large MR have aroused numerous studies owing to its potential excellent performance in the application of memory devices and spintronics.[1-3] For example, the colossal MR (CMR) was found in manganite oxides[1]; in particular, the extremely large MR (XMR) was discovered in non-magnetic compounds such as $Cd_3As_2$, $NbSb_2$, TaAs, NbP and $WTe_2$.[4-8] Among those materials, an emergent quantum behavior with nonsaturating XMR even up to the field of 60 T was found in $WTe_2$.[8] The unique feature of the XMR material is the "turn-on" (t-o) behavior of the temperature dependent resistivity $\rho(T)$. The definition of the t-o behavior can be found everywhere.[8] The discovery of XMR in $WTe_2$ has triggered extensive research to reveal the origins of XMR in this material. The reported angle-resolved photoemission spectroscopy (ARPES) results supported the proposition of an electron-hole (*e-h*) compensation mechanism, which was also supported by the quantum oscillation experiments.[9-10] Wu *et al.* found that a new type of Lifshitz transition induced by temperature is responsible for the origins of XMR in $WTe_2$.[11] At the same time, the electronic structure changes in $WTe_2$ has also been proposed to be a possible origin of the t-o effect.[12,13] However, Wang *et al.* questioned such an interpretation and demonstrated that the Kohler's rule could explain the t-o behavior of $WTe_2$.[14] So far, the origin of the t-o behavior of the XMR effect remains an open question in $WTe_2$. Very recently, orthorhombic phase $MoTe_2$ ($T_d$-$MoTe_2$) has also been reported to exhibit the XMR effect. The large and nonsaturating MR with the value of 61700% under a magnetic field of 33 T at 2 K is observed.[15, 16] Interestingly, the t-o behavior is also observed in $T_d$-$MoTe_2$. However, the quantum oscillation measurement shows that there is an obvious difference of the Fermi surfaces (FSs) between $T_d$-$MoTe_2$ and $WTe_2$.[16] That means the mechanism of the XMR in $T_d$-$MoTe_2$ may be different from that of $WTe_2$. Thus, the origins of the XMR effect in $T_d$-$MoTe_2$ are deserved to be clarified, which may be helpful for understanding the t-o effect in isostructural $WTe_2$ and other XMR materials. In this paper, we focus on the angle dependent MR, the temperature dependent thermoelectric and Hall effects to investigate the origins of the XMR effect in $T_d$-MoTe$_2$. The three dimensional (3D) character of the MR in $T_d$-MoTe$_2$ will be revealed and the origins of XMR effect will be discussed.

$T_d$-MoTe$_2$ single crystals were grown by the flux method using NaCl as solvent. Mo (Alfa Aesar, 99.9 %), Te (Alfa Aesar, 99.99 %) and NaCl (Alfa Aesar, 99.9 %) powders were ground and placed into alumina crucibles and the recipe is following the reported one.[15] The size of the obtained flake-like crystals is 2*0.5*0.02 mm$^3$ with (00$l$) plane, as shown in the inset of Fig. 1. The single crystals are air-stable and can be easily exfoliated. Powder X-ray diffraction (XRD) experiments were performed by the PANalytical X'pert diffractometer using the Cu $K_{\alpha 1}$ radiation ($\lambda$=0.15406 nm) at room temperature. The element analysis of the single crystals was performed using a commercial energy dispersive spectroscopy (EDS) microprobe. The four probe resistivity measurements were carried out and the magnetic field is along the $c$ direction and the current $I$ flows along the $a$-axis, the configuration is shown in the inset of Fig. 3(c). The magneto-thermoelectric power (TEP) $S(T)$ measurements were performed by a dc alternating temperature gradient technique. The magnetic field is along the $c$-axis and the temperature gradient $\Delta T$ flows along the $a$-axis. All the crystals were cleaved before doing the measurements. Temperature dependent Hall resistivity was measured using a five probe method. All the measurements were done in a Quantum Design Physical Properties Measurement System for (PPMS-9T and PPMS-16 T) 1.8 K<T<350 K and H<16 T. The electronic structures were calculated using the full-potential linearized augmented plane-wave (LAPW) method within the density functional theory (DFT), as implemented in the WIEN2K package.[17,18] The exchange-correlation potential was in the form of Perdew–Burke–Ernzerhof (PBE) with generalized gradient approximation (GGA). The plane-wave cutoff was determined by $R_{MT}K_{max}$=8. Self-consistent calculations were done with 4000 $k$ points in the Brillouin zone (BZ) and the total energy was converged to within 0.00001 Ry. The Fermi surfaces were calculated with 20000 $k$ points in the BZ and plotted with the program XCrySDen.[19] The spin-orbit coupling was included in the calculations.

Figure 2 shows the temperature dependent resistivity $\rho_{xx}(T)$ and the field dependent MR with the magnetic field H//$c$. The $\rho_{xx}(T)$ under different magnetic fields is presented in Fig. 2(a). With the increasing magnetic field, the $\rho_{xx}(T)$ shows the t-o effect in $T_d$-MoTe$_2$. That is, in the absence of the magnetic field, a metallic behavior is found throughout the entire temperature range. When the magnetic field $H$ is larger than 3 T, the $\rho_{xx}(T)$ shows the minimum at $T^*$ and tends to increase with decreasing temperature below $T^*$, which is called the t-o effect in $\rho_{xx}(T)$. However, the temperature of the structural transition from 1T' phase to $T_d$ one does not change even up to the field of $H$=8.5 T, as shown in the inset of Fig. 2(a), where the red arrow marks the structural transition from monoclinic phase to orthorhombic phase around $T_S$=240 K, which has been confirmed by the scanning transmission electron microscopy (STEM) result.[20-23] Figure 2(b) presents the $\rho_{xx}(T)$ as a function of the magnetic field at different temperatures. At the high temperature, the value of $\rho_{xx}(T)$ is nearly constant with increasing magnetic field. With decreasing temperature, the $\rho_{xx}(T)$ increases dramatically under the high magnetic field. Figure 2(c) displays the magnetic field dependence of the MR (defined by $(\rho_{xx}(H)-\rho_{xx}(0))/\rho_{xx}(0)$) at different temperatures. The MR at $T$=2 K is about 4000% under magnetic field of $H$=8.5 T, which is comparable with the reported one.[15] As shown in Fig. 2(c), the fitting curves according to the powder law $MR \sim H^n$ with $n$=1.9 and $n$=2.6 for $H$>1 T and $H$<1 T at $T$=30 K , 40 K and 70 K and with $n$=1.9 at low temperatures. In the low field, the $MR \sim H^{2.6}$ behavior was attributed to the contribution of the Dirac quantum states ($MR \sim H$) and the trival parabolic band ($MR \sim H^2$). In the high field, the $MR$ follow the semiclassical behavior may because the Dirac state comparatively hidden by the improved parabolic band.

To investigate the anisotropy of the resistance in $T_d$-MoTe$_2$, we performed the angle dependent magneto resistivity $\rho_{xx}(H,\theta)$ at fixed temperatures. As shown in the inset of Fig. 3(c), $\theta$ is defined as the angle between the magnetic field and the $c$-axis and the current flows along the $a$-axis. As shown in Fig. 3(a) and (b), the $\rho_{xx}(H,\theta)$ decreases with increasing angle $\theta$ from 0° to 90° at 5 K and 50 K, suggesting an anisotropy of transport behavior. Furthermore, for a 2D system, the values of $\rho_{xx}(H,90°)$ (it means $H // b$ axis) should be a constant because only the perpendicular component $H\cos\theta$ contributes to the MR.[12] In our case, the results are not consistent with above hypothesis. We found that all the $\rho_{xx}(H,\theta)$ curves obtained at a fixed temperature with different angels can be scaled into a single curve $\rho_{xx}(H,0°)$ with a field factor $\varepsilon_\theta=(cos^2\theta+\gamma^2sin^2\theta)^{1/2}$, where $\gamma$ is a constant and related to the anisotropy of FS. The $\gamma^2$ can be interpreted as the ratio of effective masses of electrons moving in directions of $\theta$=0° and 90°, i.e., along the $a$-$c$ plane and the $a$-$b$ plane for our case. [23-25] Thus the value of $\gamma^2$ can be obtained by $m_{ac}^*/m_{ab}^*$.[24] As shown in Figs. 3(c) and (d), the value of $\gamma$ changes from $\gamma$=4 for $T$=5 K to $\gamma$=1.2 for 50 K. Figure 3(e) shows that the resistivity of $T_d$-MoTe$_2$ can be well described by the following scaling relation: $\rho_{xx}(H,\theta)=\rho_{xx}(\varepsilon_\theta H)$ with $\varepsilon_\theta=(cos^2\theta+\gamma^{-2}sin^2\theta)^{1/2}$.

Let's try to understand the observed scaling behavior in $T_d$-MoTe$_2$. Actually, it has been proposed to illuminate the anisotropic properties of high temperature superconductors and graphite.[24-28] The detailed description of the scaling model can be found in Supporting Information. Very recently, Thoutam *et al.* found that the mass anisotropy of WTe$_2$ is as low as 2 at high temperature, much smaller than the expected value for a 2D system. The distortion of the tellurium layers is responsible for the small anisotropy of WTe$_2$, which is consistent with the latest quantum oscillation results.[10] For $T_d$-MoTe$_2$, the value of $\gamma$ changes from 4 at $T$=5 K to 1.2 at $T$>70 K, which is also much smaller than those of graphite and Bi$_2$Sr$_2$CaCu$_2$O$_{8-\delta}$.[27, 28] Although the model $\varepsilon_\theta=(cos^2\theta+\gamma^{-2}sin^2\theta)^{1/2}$ needs to be taken with caution due to the multiband of $T_d$-MoTe$_2$, it still can give some information on the mass anisotropy of materials.[12]

To investigate the origins of XMR in $T_d$-MoTe$_2$, we firstly focus on the t-o effect under the applied magnetic field observed in $T_d$-MoTe$_2$, which is also a distinct feature of other XMR materials.[4-8, 14] Actually, the t-o effect has been observed in graphite and bismuth and its origins are suggested to be related to the band gap opening at the temperature $T^*$.[29] However, our results of temperature dependent *MR* at different fields do not support the band gap opening mechanism. As shown in Fig. 4(a), the temperature dependent *MR* curves at the fixed field, divided by the value of *MR* at $T$=2 K, collapse into one curve. As we know, if there exists a gap opening induced by the magnetic field, the slope of the *MR* curve under higher magnetic field will be much steeper than that under lower one.[29] Therefore, the t-o effect in $T_d$-MoTe$_2$ may be not due to the metal-insulator transition. To further explore the origins of XMR in $T_d$-MoTe$_2$, the temperature dependent mass anisotropy $\gamma$ is also presented in Fig. 4(a). The temperature evolution of $\gamma$ obtained from *MR*, which is related to the anisotropy of FS, does not follow that of *MR*. The temperature interval $\Delta T$ is about 10 K. This behavior in $T_d$-MoTe$_2$ is different from the findings of WTe$_2$ that the temperature dependence of $\gamma$ matches well with that of *MR*.[12] As shown in Fig. 4(a), the *MR* increases drastically as temperature decreases below $T_{MR}^{onset}$~35 K. However, for the mass anisotropy $\gamma$, with decreasing temperature, it starts to increase at $T_\gamma^{onset}$~60 K and then rises rapidly. It means that the electronic structure changes may happen in $T_d$-MoTe$_2$ before the emergence of MR. Additionally, as shown in Fig. 4(b), $\rho_{xx}(T)$ can be well fitted by the Fermi liquid formula: $\rho_{xx}=A+BT^2$, in which $\rho_0$ and $A$ are the residual resistivity and a constant, respectively. That implies the electron-electron scattering is dominant below $T$=70 K.

To understand further the novelty of the transport properties of $T_d$-MoTe$_2$ under magnetic field, we continue with investigating the temperature dependence of thermoelectric power (TEP) $S(T)$. In general the Seebeck coefficient is related to the derivative of the density of states at the Fermi level and can provide indispensable information in understanding resonant compensated systems.[10] For example, a Lifshitz transition driven by the temperature is observed at $T$=160 K in WTe$_2$. A unique feature in the Lifshitz transition is defined as the turning point in the slope of $dS(T)/dT$.[11] In other words, a topological reconstruction of the FS induces an anomaly in the Seebeck coefficient. Figure 4(c) shows the temperature dependent TEP under different magnetic fields. $T$=240K corresponds to the structural phase transition. The sign of $S(T)$ changes from negative to positive at $T$=120K, indicating the emergence of electron pocket in the range of intermediate temperature and holes are dominated carriers at low temperature. However, the values of $S(T)$ keeps nearly unchanged when magnetic fields are applied. We can find that the turning point of slope of $dS/dT$ appears around 150 K. We also performed systematic Hall effect measurements on studied crystals because it is a useful method to give relevant information on the evolution of the electronic structure. Figs. 5(a-f) display $\rho_{xy}$ and $\rho_{xx}$ as a function of magnetic field for fixed temperature ranging from 20 to 170 K. The Hall conductivity $\sigma_{xy}=\rho_{yx}/(\rho^2_{yx}+\rho^2_{xx})$ is obtained from the original experimental data. The Hall conductivity can be well fitted by the semiclassical two-band model, which is given by:

$$\sigma_{xy} = [n_e\mu_e^2 \frac{1}{1+(\mu_e^2 B)} - n_h\mu_h^2 \frac{1}{1+(\mu_h^2 B)^2}]eB \qquad (1)$$

Where $n_e$ (or $n_h$), $\mu_e$ ($\mu_h$) represent the electron (or hole) densities and electron (or hole) mobilities. Figure 5(g) and (h) show the fitting results of $\sigma_{xy}$, in which the open symbols represent the experimental data and the solid line in the inset of Fig. 5(g) represents the fitting result at $T$= 20K. Based on the fitting results, we plotted the temperature dependence of carrier densities and carrier mobilities curves in Figs. 5(i) and 5(j). Furthermore, the ratio between the carrier densities ($n_h/n_e$) and carrier mobilities ($\mu_h/\mu_e$) are displayed in Figs. 4(d) and (e). As shown in the Fig. 4(d). Firstly, as the temperature decreases, the slight upturn of $n_h/n_e$ is around 150 K, combined with the turning point in thermo power case, indicates a possible Lifshitz transition. These observation is similar with the case in WTe$_2$[11]. Another interesting feature is the increasing of the value of $n_h/n_e$ occurs around 60 K, which is closes to the turning point marked in the temperature dependent mass anisotropy $\gamma$. We found that T$_d$-MoTe$_2$ may undergo two possible FS reconstructions: the first one is corresponding to the anomaly of $n_h/n_e$ around $T\sim150$ K, indicating that a possible Lifshitz transition exists in T$_d$-MoTe$_2$; the second one is around $T\sim60$ K, corresponding to the sharp increase of $n_h/n_e$.

To validate it, the electronic structure basis for the observed phenomena is investigated by the first-principles calculations, which is shown in Fig. 6. Here we only focus on the T$_d$ phase at low temperature, where the XMR effect is observed. Due to the spin-orbit coupling, the four bands crossing the Fermi energy split into eight ones, which are indicated in different colors in Fig. 6(a). The corresponding FS is shown in Fig. 6(c), in which the hole pockets (in yellow for the outermost one) centered at the Γ point and the electron pockets (in purple for the outermost one) located along the Γ-$X$ and Γ-$Y$ directions. As the temperature increases, the Fermi energy will be slightly elevated. In Figs. 6(d)-(f), we show the FSs when the Fermi energy moves to the three positions denoted as $E_{F1}$, $E_{F2}$, and $E_{F3}$ in Fig. 6(b), which are respectively 8.0, 17.9 and 30.0 meV higher than $E_{F0}$. We can see that as temperature increases, the topology of the electron pockets is nearly unchanged, though the size is slightly enlarged. In contrast, the hole pockets change significantly. We display all the hole pockets on the right side of the FS. The pockets marked as "I", "II", "III" and "IV" correspond to the four bands in Fig. 6(b), respectively. First, when the Fermi energy moves to $E_{F1}$, the hole pocket "I" disappears; when it moves to $E_{F2}$, the hole pocket "II" disappears and there are only two hole pockets when the Fermi energy is larger than $E_{F2}$. Second, both the topology and the size of the hole pockets change gradually as the Fermi energy moves from $E_{F0}$ to $E_{F3}$. Therefore, the temperature-induced reconstruction of the FS is indeed observed, mainly originating from the hole pockets. On the one hand, the size change of the electron and hole pockets will influence the corresponding carrier concentrations; that is, when the temperature is decreased, the hole (electron) concentration increases (decreases). When the Fermi energy moves down to $E_{F2}$ ($E_{F1}$), there will be a sharp increase in the hole concentration since the contribution coming from the band II (I) emerges. At low temperature, the $e$-$h$ compensation is achieved. On the other hand, the change of the anisotropy of $\rho_{xx}(H,\theta)$ is determined by the change of the FS topology. We evaluate the effective masses $m^*$ and the anisotropy $\gamma$ of the effective mass for the four band extrema in Fig. 6(b), which are listed in Table S1. At 0 K, all the four bands contribute to the transport properties of $T_d$-MoTe$_2$. As the temperature increases, the contribution from bands "I" and "II" gradually disappears, so the anisotropy of $\rho_{xx}(H,\theta)$ becomes smaller at higher temperature, which is consistent with our experimental results. The temperatures corresponding to the energies $E_{F1}$ and $E_{F2}$ ($E_F \sim k_BT$) are estimated to be 92 and 206 K, respectively, a little higher than the two critical temperatures where the upturn temperature of the ratio of carrier density appear (see Fig. 4(d)). The computational results indicate that the temperature-induced evolution of the electronic structure plays an important role in determining the low-temperature phenomena observed in $T_d$-MoTe$_2$.

To explore the possible origins of the XMR in $T_d$-MoTe$_2$, we first paid attention to the proposed hypotheses in some typical XMR materials. On the one hand, the *e-h* compensation effect has been widely applied to explain the XMR of WTe$_2$, TaSb$_2$, TaAs$_2$, etc.[1, 30-31] On the other hand, Tafti *et al.* [31] suggested that a particular orbital texture on the electron pocket could also be an important factor that should be considered. Such orbital texture will result in a small $\rho(0)$ due to topological protection and a large $\rho(H)$ originating from the interaction of the *d-p* mixing with the magnetic field. For $T_d$-MoTe$_2$, not only does the perfect *e-h* compensation exist below $T \sim 35$ K (shown in Fig. 4(d)), but also a mixed *d-p* orbital texture on the electron pocket that is similar to the reported result in WTe$_2$[32] is identified from the calculated band structure (see Fig. S5). Now, we focus on the value of $n_h/n_e$ as a function of the temperature of $T_d$-MoTe$_2$. As shown in Figs. 4(a) and (d), when the temperature is decreased, the value of $n_h/n_e$ starts to sharply increase around $T \sim 60$ K and tends to be equal at $T=35$ K; meanwhile the *MR* increases drastically. In other words, the perfect *e-h* compensation benefits the XMR in $T_d$-MoTe$_2$ and the electronic structure change around $T \sim 60$ K is the driving force of the *e-h* compensation. Therefore, we may confirm that the XMR in $T_d$-MoTe$_2$ is triggered by the electron structure change around $T \sim 60$ K. At the same time, we also notice that there are some generic features in WTe$_2$ and $T_d$-MoTe$_2$, such as the XMR effect, the similar temperatures of the Lifshitz transition, and the possible electronic structure change at the low temperature. Especially, the Lifshitz transition at the high temperature is likely to be a general feature of WTe$_2$ and $T_d$-MoTe$_2$; meanwhile, it may be a prodromal FS reconstruction before the appearance of the electron structure change at the low temperature ($T \sim 60$

K for $T_d$-MoTe$_2$ and $T$~50 K for WTe$_2$), which triggers the XMR of two compounds. We may expect that the Lifshitz transition is a possible necessary element for the presence of XMR in WTe$_2$ and $T_d$-MoTe$_2$. Considering the particular orbital texture, we can conclude that the XMR in $T_d$-MoTe$_2$ may originate from the combination of the perfect *e-h* compensation and a mixed *d-p* orbital texture on the electron pocket. However, more experiments at low temperatures, such as ARPES, scanning tunneling spectroscopy (STS) measurement under high magnetic field, are needed to validate it in future.

In summary, we found that $T_d$-MoTe$_2$ exhibits a moderate 3D anisotropy and the anisotropic effective mass $\gamma$ is obtained from the scaling relation $\varepsilon_\theta=(sin^2\theta+\gamma^{-2}cos^2\theta)^{1/2}$. Electronically 3D character with $\gamma$ as low as 1.9 is observed in $T_d$-MoTe$_2$. With the help of the temperature dependent TEP measurements and temperature dependent carrier density that was extracted from Hall measurements, a Lifshitz transition can be verified around $T$~150 K. Our observations also predict a possible electronic structure change around $T$~60 K. The XMR effect in $T_d$-MoTe$_2$ possibly originates from the combination of the *e-h* compensation and a particular orbital texture on the electron pocket. Our findings may provide a general scaling relation for the anisotropic *MR* and help to understand the origins of the *MR* effect in other systems, such as the Weyl semimetals and the Dirac ones.

## Acknowledgements

This work was supported by the National Key Research and Development Program under contracts (2016YFA0300404, 2016YFA0401803) and the Joint Funds of the National Natural Science Foundation of China and the Chinese Academy of Sciences' Large-Scale Scientific Facility under contract (U1432139), the National Nature Science Foundation of China under contract (11674326, 11404342), and the Nature Science Foundation of Anhui Province under contracts 1508085ME103. The authors thank Dr. Chen Sun for her assistance in editing the manuscript.

**Figure 1:**

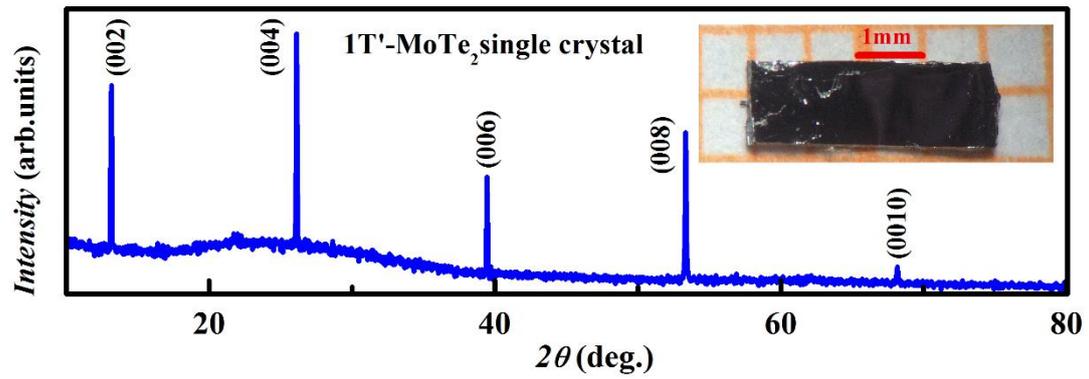

**Fig. 1.** X-ray diffraction profile of 1T'-MoTe$_2$ single crystal along the *c*-axis direction at room temperature. Inset shows the picture of the present studied single crystal. The size is approximately 2*0.5*0.02 mm$^3$.

**Figure 2:**

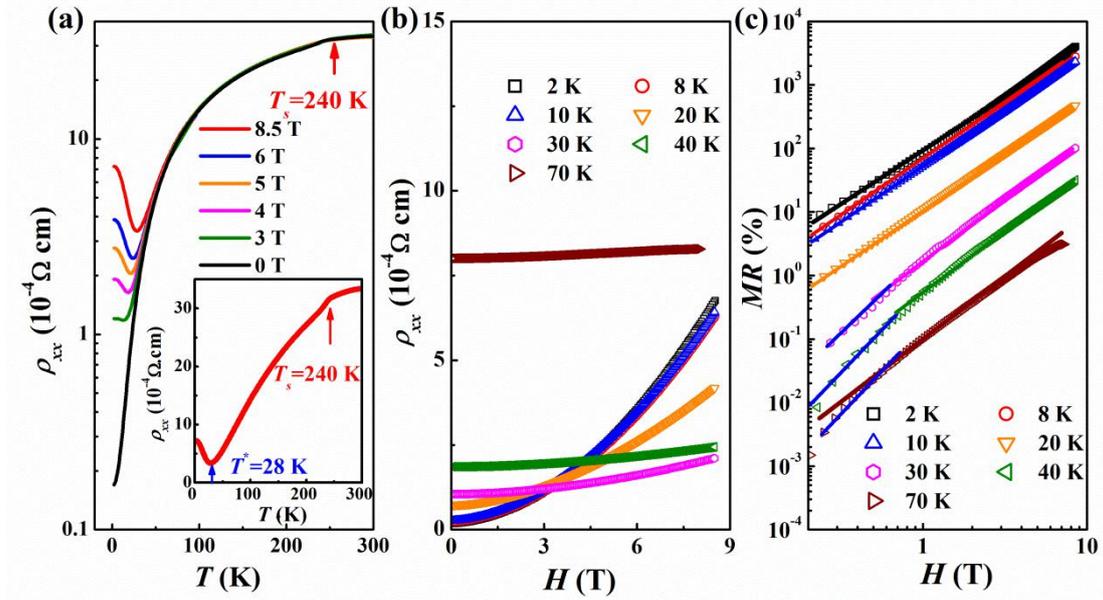

**Fig.2.** The temperature and field dependence of the XMR in $T_d$-MoTe$_2$. **(a)** The temperature dependence of the resistivity at various magnetic fields. The inset shows the $\rho_{xx}$ versus temperature at H=8.5 T with the red arrow marking the structural transition temperature $T_S$ and the blue one marking the t-o temperature $T^*$; **(b)** The evolution of the resistivity as a function of magnetic field $\rho_{xx}(H)$ at different fixed temperatures; **(c)** Power law plot of $MR \sim H^n$ at various temperatures. The solid lines are the fitting results corresponding to different temperatures.

**Figure 3:**

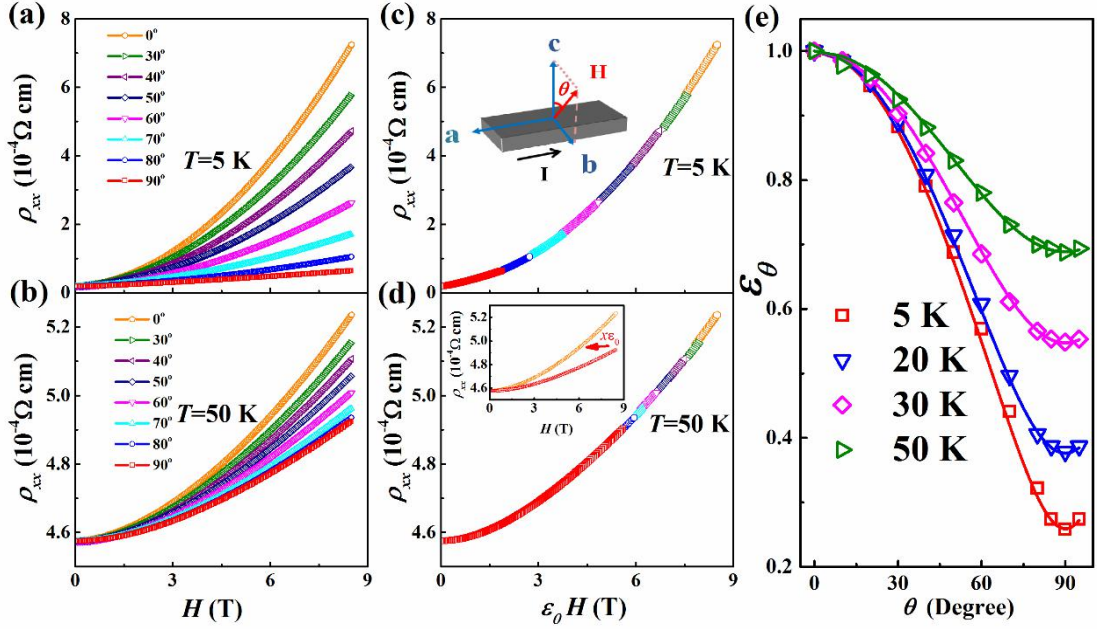

**Fig. 2.** Scaling behavior of the angle dependence of resistivity as a function of magnetic field at T=5 K and 50 K, respectively. **(a)** and **(b)**, Angular evolution of $\rho_{xx}(H)$ curves extracted at T=5K and 50 K, respectively; **(c)** and **(d)**, The measured curves at different angles in **(a)** and **(b)** collapse into one curve by using horizontal coordinate transformation with scaling factor $\varepsilon_\theta H$ substituting $H$ at T=5 K and 50 K respectively. The inset of **(c)** shows the schematic configuration of the angle dependence of MR measurement. $\theta$ is defined as the angle between the direction of the magnetic field and the *c*-axis. The inset of **(d)** shows schematically the scaling approach with the data in 90° and 0° and the red arrow indicates the scaling orientation; **(e)** The evolution of the scaling factor $\varepsilon_\theta=(sin^2\theta+\gamma^2cos^2\theta)^{1/2}$ as a function of $\theta$ at various temperatures. Solid lines are the fitting results.

**Figure 4:**

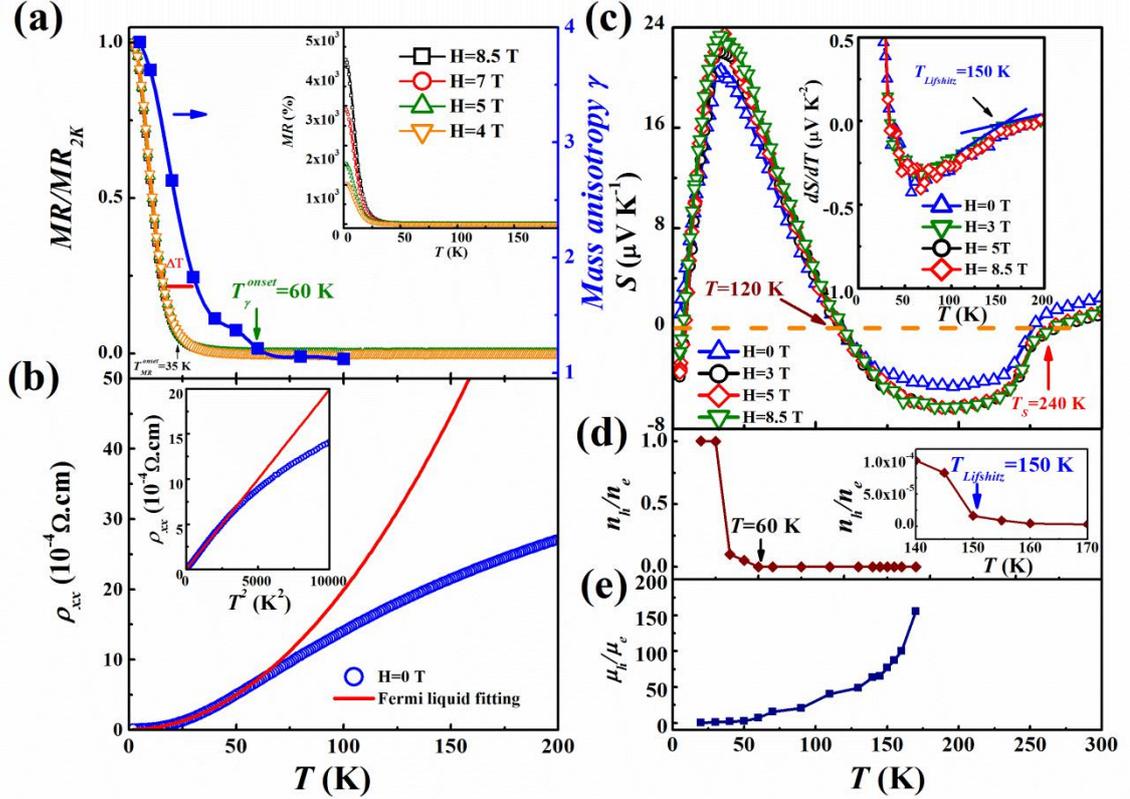

**Fig.4.** The temperature dependence of the $\rho_{xx}(T)$, MR at various fields and the mass anisotropy parameter $\gamma$. **(a)** The evolution of the normalized *MR(T)/MR(2K)* (open symbol) at various magnetic fields and the anisotropy parameter $\gamma$ (blue solid symbol) as a function of temperature. $T_\gamma^{onset}$ and $T_{MR}^{onset}$ are the onset temperatures of the XMR and the anisotropy parameter $\gamma$, respectively. Inset shows the temperature dependence of *MR* at various fields; **(b)** Temperature dependent resistivity of $T_d$-$MoTe_2$. Low temperature region is a fitting result according to the Fermi liquid formula: $\rho_{xx}=A+BT^2$. The inset shows the data plotted as a function of the $T^2$; **(c)** Temperature dependence of the Seebeck coefficient *S(T)* at various fixed magnetic fields. Orange-dashed line represents the zero value of the Seebeck coefficient. The inset shows the *dS/dT* as a function of the temperature under H=0 T and other magnetic fields. The temperatures of Lifshitz transition $T_{Lifshitz}$ are around 150 K; **(d)** Temperature dependent ratio of carrier density ($n_h/n_e$) extracted from Hall effect measurements. Inset: ($n_h/n_e$) as a function of temperature between 140K to 170K; **(e)** Temperature dependent ratio of mobility density ($u_h/u_e$) extracted from Hall effect measurements.

**Figure 5:**

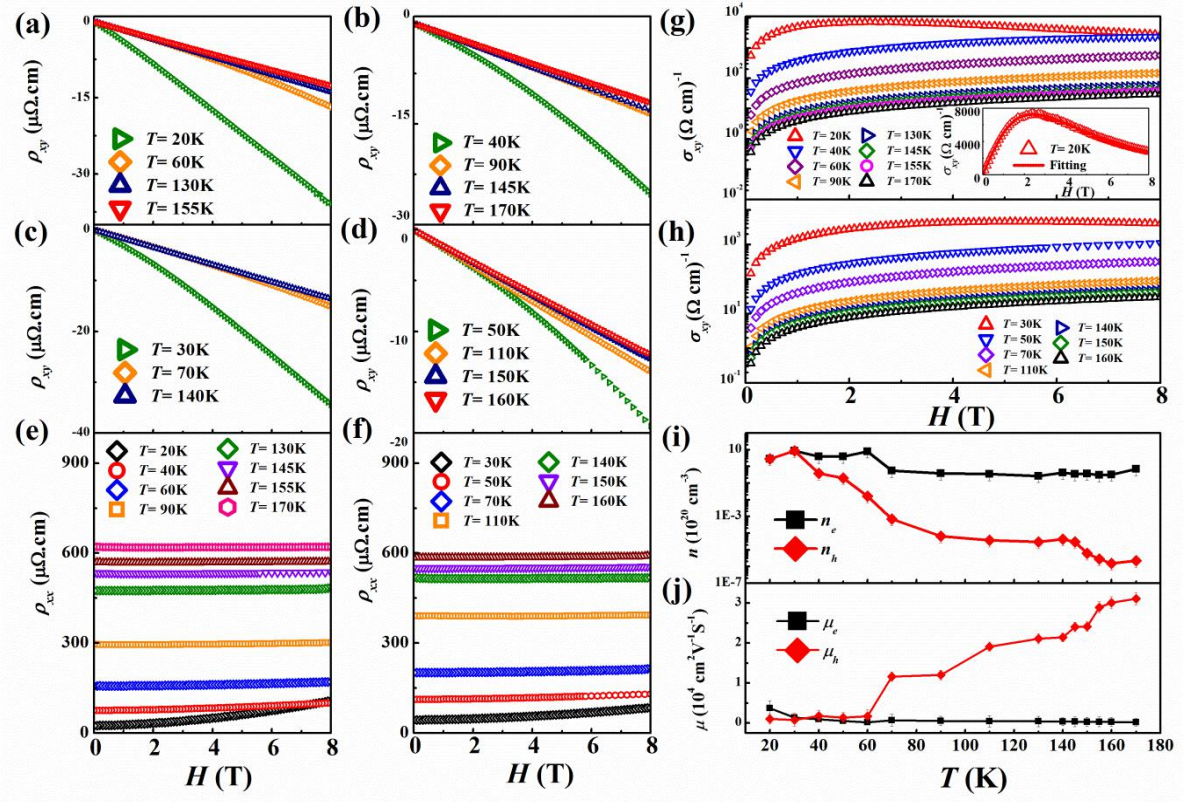

**Fig.5. (a)-(d)** The field dependence of the $\rho_{xy}(T)$ at various temperatures; **(e)-(f)** The field dependence of the $\rho_{xy}(T)$ at various temperatures; **(g)-(h)** The field dependence of the Hall conductivity $\sigma_{xy}(T)$ at various temperatures. Open symbols represent the experimental results and the solid lines represent the fitting results based on two-band model.

**Figure 6:**

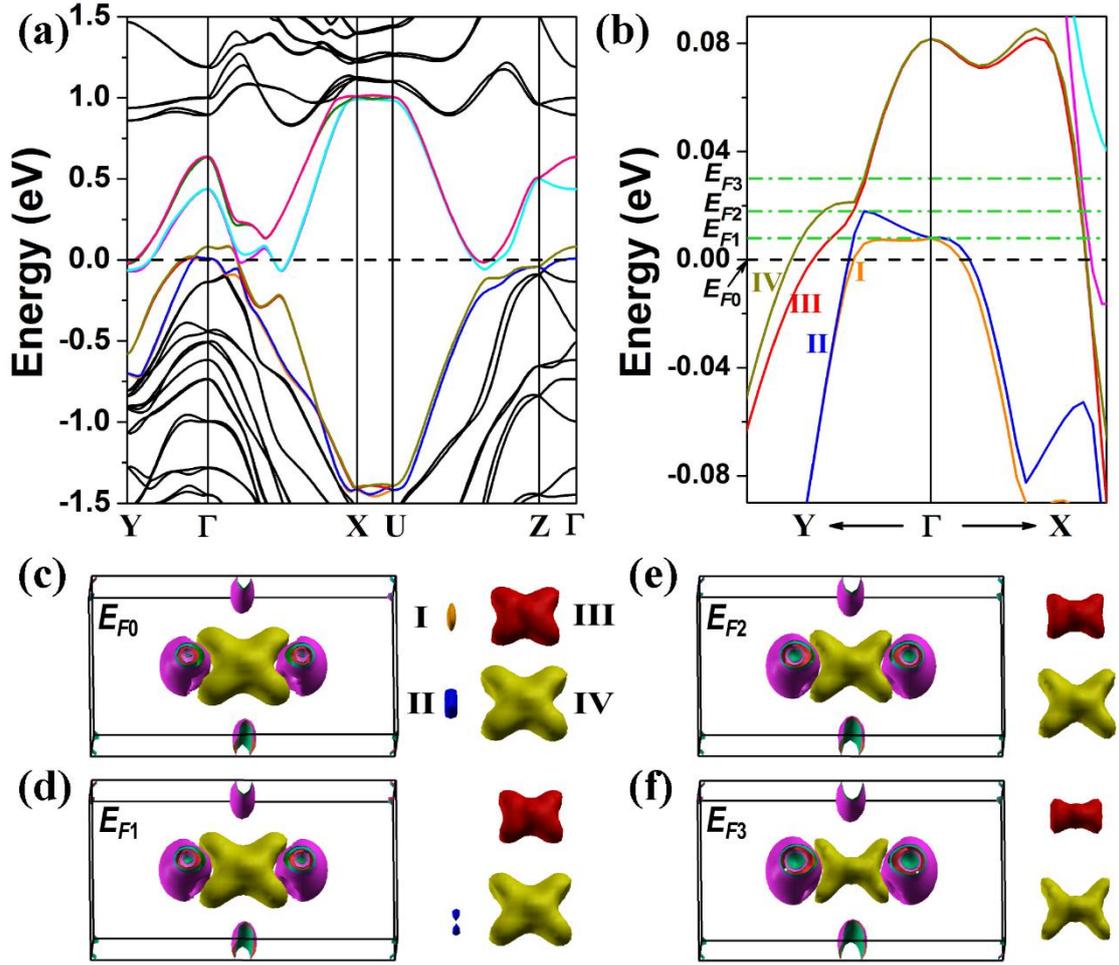

**Fig. 6.** The electronic structure of $T_d$-MoTe$_2$. **(a)** Band structure of $T_d$-MoTe$_2$; **(b)** Enlarged band structure along $Y$-$\Gamma$-$X$ direction. The four valence bands crossing the Fermi energy are denoted as "I", "II", "III" and "IV", respectively; **(c)** Fermi surface (FS) at 0 K and **(d)**, **(e)** and **(f)** are the FSs when the Fermi level moves to the energies $E_{F1}$, $E_{F2}$ and $E_{F3}$, respectively. The pockets on the right side of the FSs are the hole pockets centered at the $\Gamma$ point, with "I", "II", "III" and "IV" correspond to the four bands in **(b)**, respectively.

**Figure 7:**

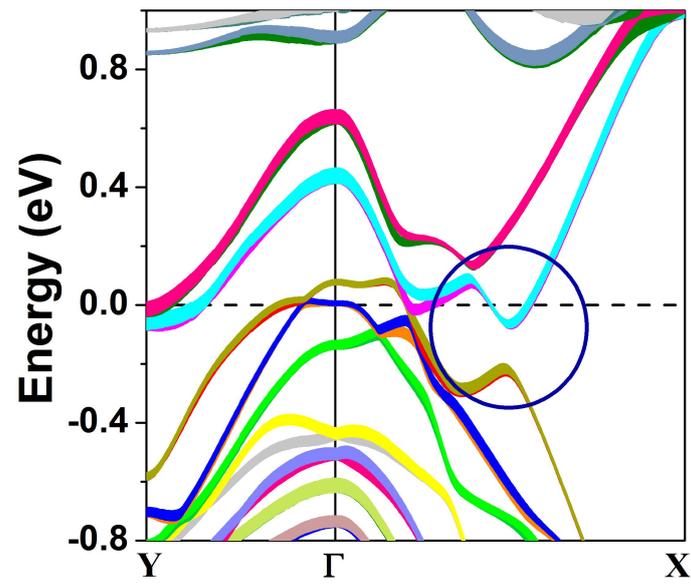

**Fig. 7.** Band structure of $T_d$-MoTe$_2$. The crossing between Mo $d$ states (thick bands) and Te $p$ states (thin bands) within the region of the blue circle gives a mixed $d$-$p$ orbital texture to the electron pocket.